\documentclass[11pt,twoside]{article}

\usepackage{asp2004}
\usepackage{epsf}
\usepackage{graphicx}
\usepackage{psfig}
\usepackage{lscape}

\newcommand{\commentpcf}[1]{}
\def\lya{\hbox{L$\alpha$}}

\def\glong{$\ell$}
\def\glat{$b$}

\def\kms{\hbox{km s$^{\rm -1}$}}
\def\deeg{\hbox{$^\circ$}}
\def\cc{\hbox{cm$^{-3}$}}
\def\cmtwo{\hbox{cm$^{-2}$}}
\def\NHI{\hbox{$N$(H$^{\rm o }$)}}

\def\nHI{\hbox{$n$(H$^{\rm o }$)}}
\def\nHeI{\hbox{$n$(He$^{\rm o }$)}}

\def\HI{\hbox{H$^{ \rm o }$}}
\def\OI{\hbox{O$^{ \rm o }$}}
\def\ArI{\hbox{Ar$^{ \rm o }$}}
\def\DI{\hbox{D$^{ \rm o }$}}

\def\HeI{\hbox{He$^{ \rm o }$}}
\def\NeI{\hbox{Ne$^{ \rm o }$}}
\def\NI{\hbox{N$^{ \rm o }$}}

\def\TiII{\hbox{Ti$^{ \rm + }$}}
\def\FeII{\hbox{Fe$^{ \rm + }$}}

\def\Vhc{\hbox{$V_\mathrm{HC}$}}

\begin{document}

\title{The Heliosphere as a Probe of Small Scale Structure}

\author{P. C. Frisch}
\affil{University of Chicago, Department of Astronomy and Astrophysics} 

\begin{abstract} The heliosphere serves as a probe of interstellar
material (ISM) close to the Sun.  Measurements of ISM inside and
outside of the heliosphere show that we reside in typical warm
partially ionized ISM that can be successfully modeled using
equilibrium photoionization models.  The heliosphere wake leaves a
$\sim 200 \times 1000$ AU trail in space of low density, $n <0.05$
\cc, cooling plasma comingled with ISM.  The closest ISM flows through
the solar vicinity at $V_\mathrm{LSR} \sim 20$ \kms, with an upwind
direction towards the Scorpius-Centaurus Association.  Clouds in this
flow have thicknesses typically $<1$ pc.  The flow is decelerating,
with velocity variations of $ \sim 3 - 4$ \kms\ pc$^{-1}$.  The
$\alpha$ Oph sightline shows evidence of a cold, possibly tiny, cloud.
\end{abstract}

\section{Introduction}

Our Sun lives in what appears to be a typical warm partially ionized
low column density interstellar cloud, yet the interstellar medium
(ISM) within a few parsecs of the Sun shows ionization, velocity, and
abundance gradients on subparsec spatial scales.  The heliosphere both
responds to, and creates, small scale structure in the ISM.  What do
we know about the small-scale structure in the surrounding cluster of
cloudlets, and what does it tell us about small-scale structure in the
ISM in general?

The cluster of local interstellar cloudlets (CLIC, d$< 30$ pc) flows
past the Sun at a heliocentric velocity of $V_\mathrm{HC} \sim 28$
\kms\ (\S \ref{sec:kinematics}).  Interstellar absorption lines
towards stars within 10 pc have \NHI$ = 0.1 - 2.0 \times 10^{18}$
\cmtwo\ \citep[for \DI/\HI$ = 1.5 \times 10^{-5}$
\cmtwo,][]{Woodetal:2005,RLII}.  The mean ISM density within 10 pc is
$<$\nHI$> \sim 0.06$ \cc.  If these clouds have the same density as
the parent cloud feeding ISM into the heliosphere, \nHI$\sim 0.18$
\cc\ \citep[][SF02,SF06]{SlavinFrisch:2002,SlavinFrisch:2006}, then
cloud thicknesses are $\sim 0.22 - 3.1$ pc.  If the width of the
observed components is formed by a mass-dependent thermal component
and a mass-independent turbulent component, then cloud temperatures
range between 1700 K and 12,600 K \citep{RLIII}.

The success of equilibrium photoionization models (SF02,SF06) in
predicting the properties of the Local Interstellar Cloud (LIC)
surrounding the Sun suggests that clouds within 10 pc may be in
pressure equilibrium with each other and with the LIC, and have
similar ionization levels (\S \ref{sec:heliosphere}).  In this case,
the cloud densities are \nHI$= 0.13 - 0.97$ \cc, the cloud thicknesses
are $0.06 - 2.2 $ pc, and the components fill 7\% to 71\% of the
individual sightlines.  The highest filling factor would then be
towards $\alpha$ Aql (5 pc), $\sim 80^\circ$ from the LSR upwind
direction of the CLIC (\S \ref{sec:kinematics}), which is consistent
with a CLIC affiliation with an expanding evolved superbubble shell
(\S \ref{sec:kinematics}).

The heliosphere boundary conditions vary over time-scales of $10^3 -
10^5$ years because of the velocity and ionization gradients of the
CLIC (\S \ref{sec:kinematics},\ref{sec:heliosphere}).  The Sun appears
to have entered the CLIC within the past $\sim 100,000$ years,
influencing the galactic cosmic ray (GCR) flux at the Earth
\citep[][FS06]{FrischSlavin:2006book,FrischSlavin:2006astra}.  The
paleoclimate record of short-lived radioisotopes resulting from GCR
temporal variations is an in situ probe of small scale ISM structure.
Interstellar dust and gas properties inside of the heliosphere depend
on the heliosphere boundary conditions.  Large grains and neutrals
from the surrounding Local Interstellar Cloud (LIC) penetrate and
interact with the solar wind plasma (\S \ref{sec:heliosphere}).

\section{Solar Wake:  Tiny Clump of Cooling Plasma and Hot Neutrals}

The interaction between ISM and the heliosphere must generate
small-scale structure in the ISM because the relative Sun-cloud
motion, $\sim$26.4 \kms, generates a wake of comingled ISM and cooling
solar wind that trails the Sun through space.  Our Sun looses $\sim 2
\times 10^{-14} ~M_\mathrm{sun}$ per year.  Voyager 1, which is now
exiting the heliosphere in a direction roughly towards BD+12 3139,
detected the solar wind termination shock \footnote{The termination
shock of the solar wind, where the solar wind becomes subsonic, is
observed as a low energy flux of ``termination shock particles''
(protons), and the factor of $\sim 3$ compression of the solar wind
magnetic field \citep{Stoneetal:2005,Burlagaetal:2005}.}  at 95 AU in
a direction offset $\sim 30$\deeg\ away from the heliosphere nose and
34.7\deeg\ above the ecliptic plane.  The relative velocity of the Sun
and surrounding interstellar cloud is 26.4 \kms\ \citep{Witte:2004},
so that for a solar wind density of $\sim$6 particles \cc\ at 1 AU,
the solar wind leaves a trail in space of cooling solar wind material
with average density $\sim 0.05 $ \cc.  The length of this solar wake,
$\sim 10^3$ AU, is limited by the reconnection rate in the solar wind
and charge-exchange between solar wind protons and interstellar H
\citep{Yu:1974}.  The $\sim 200 \times 10^3$ AU solar wake has an
aspect ratio of $\sim 1/5$.  Random sightlines through the solar wake
will yield column densities of energetic neutral hydrogen atoms
(ENAs), created by charge exchange between the solar wind and low
velocity interstellar H-atoms, of \NHI$< 8 \times 10^{14}$ \cmtwo.
The full-width-half-max of these features should be less than twice
the solar wind speed, or generally $<500$ \kms\ for slow solar wind
with higher flux levels.  Such features could be a significant
contaminant of the damping wings of the \lya\ line towards other
stars.  Similar features should form around other cool stars embedded
in the low densities of the Local Bubble.

\section{Kinematics of ISM Flow Past Sun} \label{sec:kinematics}

\begin{figure}[ht!]
    \begin{center}
   	\includegraphics*[bb = 16 148 583 443,width=3.8in]{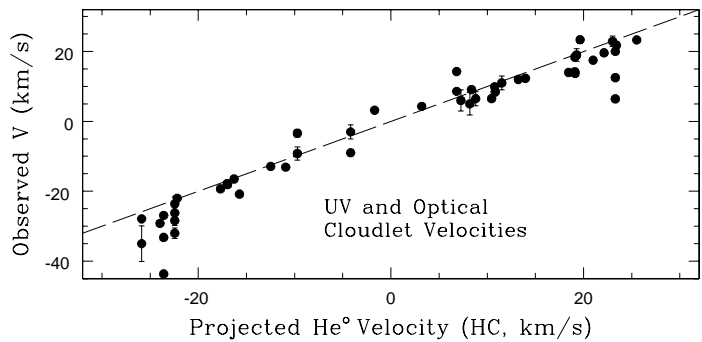}
   \end{center}
  \caption{ The observed heliocentric velocity (y-axis) of cloudlets making up the CLIC are plotted in the rest frame of the LIC (x-axis),
 which is defined by measurements of interstellar \HeI\ inside of the
 heliosphere (\S \ref{sec:heliosphere}).  The components displayed in
 this figure are observed towards stars within 30 pc of the Sun
 \citep[see][for star list]{FGW:2002}.  Note that the CLIC flow is
 decelerating.  }
\label{fig:velocity} 
\end{figure}

It was recognized long ago that the ISM inside of the solar system is
part of an outflow of ISM from the Scorpius-Centaurus Association
\citep[SCA,][]{Frisch:1981}.  In the local standard of rest (LSR),
\footnote{We follow the practice of radio astronomers and use the
Standard solar apex motion to calculate CLIC LSR velocities; see FS6
for values based on Hipparcos data).}  the bulk flow of the CLIC
corresponds to an upwind velocity vector $V_\mathrm{LSR} = -19.5 \pm
4.5 $ \kms, from the direction \glong=331\deeg, \glat=--5\deeg
\citep{FGW:2002}.  Ultraviolet (UV) and optical absorption lines trace
the radial velocities of CLIC components, and indicate the flow is
decelerating by $\sim 30 - 40$ \kms\ over the nearest $\pm 5$ pc,
averaging to $ \sim 3 - 4$ \kms\ pc$^{-1}$.  The most negative
velocity components in the upwind direction (\Vhc$\sim - 35
~\mathrm{to} ~ - 40 $ \kms) approach the Sun more rapidly than the
most positive velocity components in the downwind direction (
\Vhc$\sim + 20$ \kms) recede (Fig. \ref{fig:velocity}).  The velocity
dispersion of $\pm 4.5$ \kms\ arises partly because several individual
cloudlets are identified in this flow, with LSR velocities of --20.7
\kms\ to --24.5 \kms\ and similar upwind directions (within 25\deeg\
of each other, for Standard solar apex motion).  The extremes of the
observed velocity range are seen in stars within 5 pc of the Sun,
including $\alpha$ Aql (5 pc) in the upwind direction and $\alpha$ CMa
(2.7 pc) in the downwind direction
\citep[e.g.][]{Lallementetal:1986,FGW:2002}.

One nearby cloudlet is the Apex cloud (in the direction of solar apex
motion), which is seen towards $\alpha$ Aql and $\alpha$ Oph (14 pc)
with a density $n >$5 \cc\ if ionization is uniform
\citep{Frisch:2003apex}.  The Apex cloud also provides the only
evidence for possible cold ISM close to the Sun.  An \HI\ 21-cm
absorption feature at the Apex Cloud velocity ($V_\mathrm{LSR} = - 14
$ \kms) is seen towards the radio source NVSS J1732+125, 41 arcmin
(0.1 pc) from the $\alpha$ Oph sightline, with spin temperature
$T_\mathrm{spin} = 40$\deeg K and \NHI$ = 10^{19.70}$ \cmtwo\
\citep{Mohanetal:2001}.  Although the radio component may be beyond
$\alpha$ Oph, the \TiII\ absorption at the same velocity gives
$N$(\TiII)/\NHI$ \sim 10^{-9.6}$, which is within the uncertainties of
similar values found for higher column density gas (Frisch et al., in
preparation).  The cloud length is $\sim 0.16$ pc for $n \sim 100 $
\cc.

The physical basis of the CLIC kinematics is unknown.  Possible
processes include that the Sun is in a fragment of the SCA superbubble
shell \citep{Frisch:1995}, that the flow results from a
Rayleigh-Taylor instability in the interaction region between the Loop
I supernova remnant and Local Bubble \citep{Breitschwerdtetal:2000},
and that local ISM is caused by magnetic tension which has caused the
detachment of a magnetic flux-tube from the Loop I bubble
\citep{CoxHelenius:2003}.

The warm partially ionized (WPIM) CLIC appears similar to warm neutral
material (WNM) detected by \HI\ 21-cm observations \citep{HTII}, where
WNM temperatures range up to $\sim 5000$ K and $\sim$25\% of the \HI\
mass appears at velocities $|V_\mathrm{LSR}|>10$ \kms.  The LIC
heliocentric velocity measured from the detection of interstellar
\HeI\ inside of the solar system is --26.3 \kms\
\citep[][]{Witte:2004}, which should not be surprising since
encounters between the Sun and ISM with a relatively high Sun-cloud
velocities are more likely.

\section{Ionization:  The Heliosphere Vantage Point} \label{sec:heliosphere}

Observations of ISM inside of the heliosphere provides a unique
opportunity to constrain the physical properties of the ISM at a
single location in space using radiative transfer models to
reconstruct ionization gradients that are present in WPIM.  Generally,
absorption line data sample ISM velocity structure blended over large
distances, masking small scale structure related to ionization
gradients.  Elements with first ionization potentials (FIP) $ \ge
13.6$ eV are partially ionized since an opacity of $\tau \sim 1$ for H
and He ionizing photons is achieved for \NHI$\sim 10^{17.3}$ \cmtwo\
and \NHI$\sim 10^{17.7}$ \cmtwo, respectively.  The resulting
ionization gradients, including for \HI/\HeI, requires radiative
transfer models to recover cloud properties at any single location
(SF02,SF06).  Observations of interstellar neutrals and their
byproducts inside of the heliosphere diagnose the ionization levels.
Among partially ionized elements detected in the heliosphere are He,
O, N, Ne, and Ar.

\begin{figure}[t]
  \vspace*{2mm}
   \begin{center}
 	\includegraphics*[angle=0,width=3.2in]{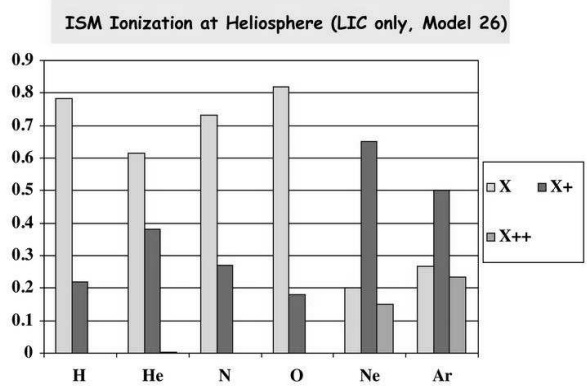}
  \end{center}
  \caption{Ionization levels of the LIC WPIM at the Sun, based on Model 26
in the LIC-only radiative transfer models of SF06.  For this model at
the solar location, \nHeI=0.015 \cc, \nHI$\sim 0.019$ \cc, $n$(e)$\sim
0.06$ \cc, and the fractional ionizations of H and He are $\sim 0.22$
and $\sim 0.38$, respectively.  Other models with slightly different
values are also acceptable (SF06).  } \label{fig:ionization}
\end{figure}

ISM types measured inside of the heliosphere include large dust grains
(radii $> 0.15 ~ \mu$m), \HeI\ atoms
\citep[e.g.][]{Witte:2004}, measurements of solar \lya\ and 584 A
florescence from \HI\ and \HeI, and measurements of H, He, O, N,
Ne, and Ar ``pickup ions'' and ``anomalous'' cosmic rays \citep[PUI,
ACR,
e.g.][]{CummingsStone:2002,Moebiusetal:2004,GloecklerGeiss:2004}.\footnote{PUIs
are formed by charge exchange between interstellar neutrals and the
solar wind.  These low energy ions are captured by the solar wind and
convected outwards, where they are accelerated beyond the termination
shock to low cosmic ray energies, $< 500$ MeV/nucleon, forming ACRs.}
Densities at the termination shock of the parent neutral interstellar
atoms can be derived from PUI and ACR data, after correction for
ionization and propagation processes.  These densities are corrected
to interstellar values through calculations of possible loss in the
heliosheath due to charge exchange or electron impact ionization
processes, termed ``filtration''.  Fig. \ref{fig:ionization} shows the
relative ionizations of the LIC, which forms the PUI parent population.

Interstellar \HeI\ at the Sun has density \nHeI$ =0.015$ \cc,
temperature $T = 6400 $ K, and heliocentric velocity --26.3 \kms (Witte 2004).  The
observed \nHeI\ value serves as a strong constraint on the SF02 and
SF06 radiative transfer photoionization models that predict the ISM
properties close to the Sun.  However, \HI\ data are less useful because
up to 50\% of interstellar H is
lost by charge-exchange in the heliosheath regions, and \HI\
trajectories depend on 
$\beta = \frac{F_\mathrm{Radiation}}{F_\mathrm{Gravitation}}$, which
is solar cycle dependent ($> 1$ during 
solar maximum and $\sim 1$ during minimum).

\begin{table}[h]
\label{tab1}
\caption{Comparisons between ISM at the Sun and other nearby PWIM}
\begin{center}
{\small
\begin{tabular}{lccc}
\noalign{\smallskip}
Ratio & Local & PUI & ISM at Sun \\
      & ISM & at the TS$^1$ &  Model 26, SF06\\
      & ($N$(X) ratios) & ($n$(X) ratios) &  ($n$(X) ratios) \\
\noalign{\smallskip}
\tableline
\HI/\HeI & $12.8 \pm 1.4$$^2$ & 6.8 (9.9)$^3$ & 12.4 \\
\OI/\NI & 8.7$^4$  & 8.8 & 7.3 \\
\OI/\HI & 0.00040$^4$ & 0.00047 & 0.00033 \\
\ArI/\OI &  0.0039 or $<0.0016  ^{4,5}$ & 0.0034 & 0.0026$^6$ \\
\NeI/\HeI   & $-$ & 0.00041 & 0.00040$^7$\\
\tableline
\end{tabular}
}
\end{center}
$^1$ {The H, O, and N PUI $n$(X) (\cc) values at the termination shock
(TS) should be corrected by $\sim 1.45, ~ 1.32,$ and $\sim 1.22$,
respectively, to recover interstellar values because of filtration
losses in heliosheath regions.}  $^2$ {Based on column density
($N$(X), \cmtwo) data for five white dwarf stars 50--79 pc from the
Sun \citep[e.g.][]{Wolffetal:1999}.}  $^3$ {The ratio is parentheses
is derived from observations of the \HI\ \lya\ backscatter
measurements inside of the heliosphere.}  $^4$ {Column densities
($N$(X), \cmtwo) from \citet{Lehneretal:2003} for stars within 70 pc.}
$^5$ {The observed \ArI/\OI\ distribution appears to be bimodal, with
small values representing highly ionized gas.}  $^6$ {This value is
based on an abundance Ar/H=2.82 ppm.}  $^7$ {This value is based on an
abundance Ne/H=123 ppm.}
\end{table}

Table 1 compares interstellar element ratios found from column density
data ($N$, \cmtwo) towards nearby stars (column 2), volume density
($n$, \cc) ratios at the solar wind termination shock, with
predictions of the RT Model 26 from SF06.  Once filtration corrections
are applied to the PUI H, N, and O data, all three sets of data are in
good agreement indicating that ISM at the Sun has similar abundances
and ionizations as other ISM within 50--70 pc.  The exception is the
sightline towards HD 149499B, 37 pc away in the LSR upwind direction
(\glong,\glat=330\deeg,--7\deeg), which is $\sim 67$\% ionized and may
trace an extended nearby HII region.  Enhanced $N$(\FeII)/$N$(\DI)
ratios in the LSR upwind direction also suggest a local ionization
source (FS06).  Otherwise, the LIC ionization level is heavily
dominated by the primary ionization sources in the third and fourth
galactic quadrants (\glong=180\deeg --270\deeg --360\deeg),
corresponding to regions of the Local Bubble within $\sim 150$ pc that
have low average interstellar opacities for $\lambda < 1500$ A
(Frisch, in preparation).

\acknowledgements 
P. Frisch would like to thank Jon Slavin for helpful
conversations.  This research is supported by NASA grants NAG5-13107
and NNG05GD36G.


\end{document}